# 50 years since the Marr, Ito, and Albus models of the cerebellum

Mitsuo Kawato[1,2,#], Shogo Ohmae[3,#], Huu Hoang[1], Terry Sanger[1,4]

[1] Brain Information Communication Research Group, Advanced Telecommunications Research Institutes International (ATR), Hikaridai 2-2-2, "Keihanna Science City", Kyoto 619-0288, Japan
[2] Center for Advanced Intelligence Project (AIP), RIKEN, Nihonbashi Mitsui Building, 1-4-1 Nihonbashi, Chuo-ku, Tokyo 103-0027, Japan
[3] Department of Neuroscience, Baylor College of Medicine, Houston, Texas, USA
[4] Departments of Biomedical Engineering, Neurology and Biokinesiology, University of Southern California, 1042 Downey Way, DRB 140, Los Angeles, California 90089, USA

# These two authors contributed equally to this work.

Correspondence to kawato@atr.jp or tsanger@usc.edu

**Abstract**

Fifty years have passed since David Marr, Masao Ito, and James Albus proposed seminal models of cerebellar functions. These models share the essential concept that parallel-fiber-Purkinje-cell synapses undergo plastic changes, guided by climbing-fiber activities during sensorimotor learning. However, they differ in several important respects, including holistic versus complementary roles of the cerebellum, pattern recognition versus control as computational objectives, potentiation versus depression of synaptic plasticity, teaching signals versus error signals transmitted by climbing-fibers, sparse expansion coding by granule cells, and cerebellar internal models. In this review, we evaluate different features of the three models based on recent computational and experimental studies. While acknowledging that the three models have greatly advanced our understanding of cerebellar control mechanisms in eye movements and classical conditioning, we propose a new direction for computational frameworks of the cerebellum, that is, hierarchical reinforcement learning with multiple internal models.

# Introduction

Masao Ito (1928-2018) made unique and profound contributions to our understanding of cerebellar functions, not only through experimental work (inhibitory action of Purkinje cells: Ito and Yoshida, 1964; Ito et al., 1964; Obata et al., 1967; adaptation of vestibulo-ocular reflex: Ito et al., 1982a; long-term depression of parallel-fiber-Purkinje-cell synapses: Ito et al. 1982b; Ito and Kano, 1982; molecular mechanisms of synaptic plasticity: Ito, 2001), but also through theoretical studies (Ito, 1970, 1974, 1984, 2006, 2008, 2011). Fifty years ago, David Marr (1969), Masao Ito (1970), and James Albus (1971) proposed computational models of cerebellar learning functions, based on neural circuits of the cerebellum (Eccles et al., 1967). As Ito acknowledged (Ito, 1970, 1974, 1984), his model was deeply influenced by Marr's model. However, the three models differed in several important respects

With recent advances in experimental and computational studies, we compare and evaluate differences between these three seminal models (See Tyrrell and Willshaw, 1992; Sanger et al., 2020 for related efforts). In particular, we extend detailed discussions of three important differences of the models: parallel-fiber-Purkinje-cell synaptic plasticity, codon theory, and internal models. For oculomotor control and classical conditioning, Ito's model was sufficiently robust to influence experimental and computational studies for the past fifty years. While we acknowledge that the three models were among the most successful and influential computational models in neuroscience, they were insufficient as a computational framework for understanding several cerebellar functions, including whole body movements (Morton and Bastian 2007; Hoogland et al., 2015; Machado et al., 2015) or cognitive functions (Ito 2008; Strick et al., 2009; Lu et al., 2012; Schmahmann et al., 2019). In conclusion, we propose a new computational framework of hierarchical reinforcement learning with multiple internal models for phylogenetically newer parts of the cerebellum and newer functions.

# Comparisons of the models of Marr, Ito, and Albus

In this section we compare several characteristics of the three models proposed by Marr, Ito, and Albus (see Table 1). First, the three models all propose that modification of parallel-fiber-to-Purkinje-cell synapses guided by climbing-fiber inputs is central to sensorimotor learning in the cerebellum (Table 1). Ito acknowledged that his model followed Marr's model regarding this fundamental principle (Ito, 1970, 1974, 1984).

|  | Marr | Ito | Albus |
| --- | --- | --- | --- |
| (1) Modification of parallel-fiber synapses on Purkinje cells conditioned by climbing-fiber inputs is central to learning. | ✓ | ✓ | ✓ |
| (2) Holistic or Complementary role of cerebellum in motor control? | Holistic | Complementary | Holistic |
| (3) Pattern recognition versus Control | Pattern recognition | Control | Pattern recognition |
| (4) Synaptic plasticity of parallel-fiber-Purkinje-cell synapses when climbing fiber is activated | LTP | LTD | Anti-Hebbian |
| (5) Does climbing fiber input represent teaching signal, error signal or unconditioned stimulus? | Teaching signal | Error signal | Unconditioned stimulus |
| (6) Sparse expansion coding by granule cells or "Codon" theory | ✓ | — | ✓ |
| (7) Internal Models | — | ✓ | ? |

**Table 1. Comparisons of the models of Marr, Ito and Albus.** A check mark indicates an included concept, – denotes not mentioned, and a question mark means that the model was ambiguous on that point.

Neural circuits of the cerebellar cortex are much more uniform, regardless of different regions and zones (Sillitoe and Joyner 2007; Zhou et al., 2014; Tsutsumi et al., 2015, 2019), than those of the cerebral cortex. Based on this hardware uniformity, researchers, and especially theorists, are tempted to assume a uniform computational theory for the cerebellum (Apps and Garwicz, 2005; Apps and Hawkes, 2009). This assumption is not well substantiated for the following reasons. One of the outstanding contributions of David Marr (1982) was his proposal of three levels of brain research: computational theory, representation and algorithm, and hardware implementation. If we accept that these three levels are somewhat independent, then hardware uniformity does not imply unified computational theory or uniform algorithms used in the cerebellum. Furthermore, different regions of the cerebellum receive mossy-fiber and climbing-fiber inputs from many different areas of the brain and the rest of the nervous system. Thus, input representations cannot be uniform either. Different regions of the cerebellum are involved in adaptation of oculomotor control, learning in conditioning, whole body movement, and cognitive functions. Correspondingly, mossy and climbing-fiber inputs range from direct inputs from sensory organs to inputs originating from higher association cortices of the cerebrum. Deep cerebellar nucleus outputs range from premotor neurons to higher cerebral association cortices via the thalamus. It would be very surprising if a single computational theory and a single representation and algorithm could explain all these diverse cerebellar functions, inputs, and outputs. Still, if we interpret the common proposal by Marr, Ito, and Albus as a computational algorithm used in the local circuits around Purkinje cells, some common algorithms and representations might be omnipresent in the cerebellum. We will explore mainly this possibility in this review.

*Holistic versus modulating control*
Marr and Albus proposed their models in a framework in which the cerebellum controls movements entirely by itself (holistic controlling role). Their models did not consider computational roles played by other brain regions or other parts of the nervous system, except for climbing-fiber inputs. In contrast, Ito conceptualized the cerebellum in a network comprising other brain regions or other parts of the nervous system (complementary, modulating role, Table 1). For oculomotor control, conditioning, and higher cognitive functions, the cerebellum works as a side path in parallel to a main path formed by other brain regions or other parts of the nervous system.

*Pattern recognition versus regression*
The major objectives of most machine learning algorithms are either classification or regression. In the simplest classification, an algorithm determines whether a given sample belongs to a given class. In a regression, an algorithm approximates a target function using output computed from multiple inputs. In many cases of regression, the target function is smooth and nonlinear. Deep neural networks can classify millions of images into thousands of object categories, in what constitutes a pattern recognition problem. Three-layer neural networks can also approximate nonlinear inverse dynamics of multi-link robotic manipulators (Miyamoto et al., 1988), a regression problem that can be used for movement control.

Marr and Albus viewed pattern recognition tasks as objectives of cerebellar learning. In contrast, Ito targeted control tasks as objectives of cerebellar learning, thus treating them as regression problems. Adaptation of oculomotor control is best treated as a learning control and regression problem. Learning in a visually guided arm-reaching task likewise can best be treated as learning control and regression problems. In learning of conditioned responses, both pattern recognition views and control views are still advocated. Khilkevich et al. (2018) reported that over-trained rabbits with eyeblink conditioning either generate conditioned eyelid responses with a stereotyped amplitude, or else fail completely to generate eyelid responses, in an all-or-nothing manner. Since this pattern of response was maintained even when noise was added to mossy-fiber input, they regard this as experimental evidence that the cerebellum is able to recognize patterns whether mossy-fiber input is a conditioned stimulus or not. Thus, conceptually, the cerebellum is often interpreted as solving the pattern recognition problem of whether mossy-fiber inputs are conditioned stimuli. However, others still view the cerebellar role in conditioned learning as learning control, including amplitudes of conditioned responses, just as the terminology changed from "eyeblink conditioning" to "eyelid conditioning" (Medina et al., 2001, 2002). In summary, most sensorimotor learning problems are better viewed as control and regression problems, rather than pattern recognition and classification problems, because many experimental data demonstrate kinematic and kinetic representations of movements in the cerebellum. We must await future studies to see how much pattern recognition and how many regression characteristics are involved in the newest functions of cerebellar higher cognitive functions.

# Synaptic plasticity in achieving supervised learning

*Sites of synaptic plasticity for cerebellar learning*
The three models assumed markedly different synaptic plasticity rules at parallel-fiber-to-Purkinje-cell synapses (Table 1). Marr assumed potentiation of these synapses when climbing-fibers, parallel fibers, and Purkinje cells are activated. Marr also postulated that in the cerebellum, only these synapses exhibit plasticity. In the past three decades, many different types of synapse have proven to be plastic (Table 2) (Hansel et al., 2001; Zhang and Linden 2006; Hirano 2013; D'Angelo et al., 2016), and many of them are Hebbian or anti-Hebbian (Gao et al., 2012). Among these, in addition to plasticity of molecular layer inhibitory interneurons, plasticity of mossy-fiber-to-deep-cerebellar-nucleus synapses has attracted experimental and theoretical interest intended to reveal their contributions to motor learning and its mechanisms (Miles and Lisberger, 1981; Lisberger and Sejnowski, 1992; Attwell et al., 2002; Tabata et al., 2002). Recently, memory transfer from parallel-fiber-to-Purkinje-cell synapses to mossy-fiber-to-deep-cerebellar-nucleus synapses has been intensively studied (Shutoh et al., 2006; Anzai et al., 2010; Yamazaki et al., 2015; Lee et al., 2015).

|  | Cell number | Synapse number |
|---|---|---|
| Parallel fiber - Purkinje | 15 million | 10,000 billion |
| Mossy fiber - Granule cells | 50 billion | 200 billion |
| Mossy fiber - Golgi cells | 5 million | 5 billion |
| Mossy fiber – deep cerebellar nucleus | 500 thousands | 5 billion |
| Purkinje cell – deep cerebellar nucleus | 500 thousands | 5 billion |

**Table 2. Numbers of plastic synapses in the human cerebellum.** The cell number column indicates the number of postsynaptic neurons. These numbers are based on Ito (1984), and Fukutani et al. (1992) and Andersen et al. (2004) for deep cerebellar nucleus.

The number of parallel-fiber-to-Purkinje-cell synapses is 50 times larger than the sum of all other plastic synapses (Table 2). The capacity of the learning system, e.g., the allowable number of non-spurious memories in a recurrent artificial neural network (Hopfield 1982; Gardner and Derrida, 1988), and approximation capacity of multi-layer perceptrons (Cybenko 1989; Funahashi 1989), increases as the number of parameters (Cover, 1965), or the number of synapses in biological networks, increases. Thus, learning capacity of the cerebellum largely depends on parallel-fiber-to-Purkinje-cell synapses, and it is reasonable that the three models emphasized these synapses so heavily. A huge learning system with trillions of learning parameters possesses immense learning capacity, but necessitates a huge learning sample for training, in order to guarantee good generalization in circumstances other than those in which the original learning occurred (Watanabe 2009, Zhang, 2017; Suzuki, 2018; Schmidt-Hieber, 2017; Amari, 2020). A smaller learning system with a small number of synapses has a small learning capacity, but is better at generalization when provided with even a small number of learning samples. Transfer learning from parallel-fiber-Purkinje-cell synapses to mossy-fiber-deep-cerebellar-nucleus synapses may benefit from these respective advantages and disadvantages of huge and small learning systems. For example, long-term memory should possess more stringent generalization capability compared with short-term memory, because the former faces larger numbers of new different situations than the latter. Consequently, a small capacity system is more suited for long-term memory in situations in which only a fixed number of training samples is given (see Sanger et al., 2020 for related discussions).

### *LTD-LTP in parallel-fiber-Purkinje-cell synapses as supervised learning machinery*

Ito proposed that climbing-fiber inputs represent error signals and that parallel-fiber-to-Purkinje-cell synapses undergo long-term depression (LTD) when both parallel-fiber and climbing-fiber inputs are activated (Table 1). Climbing-fiber inputs represent error signals originating from retinal slips in oculomotor control adaptation, including vestibulo-ocular reflexes, optokinetic responses, and ocular following responses (Simpson and Alley, 1974; Simpson et al., 1996; Kawano et al., 1996a, 1996b; Kobayashi et al. 1998; Kawano 1999). Especially for ocular following responses, parallel-fiber inputs represent large visual-field motion in visual coordinates, while climbing-fiber inputs represent motor errors derived from retinal slips in motor coordinates (Kawano et al., 1996a, 1996b; Kawato, 1999). This is a very rare example of an experimental paradigm for which sensory and motor coordinates are clearly separated. In visually guided arm reaching, climbing-fiber inputs represent errors (Kitazawa et al., 1998). Albus assumed that climbing-fiber inputs represent unconditioned stimuli, and that parallel-fiber-to-Purkinje-cell synapses obey the anti-Hebbian rule: synaptic efficacy decreases when parallel fibers, Purkinje cells, and climbing-fibers are all excited. In classical conditioning, climbing-fiber inputs can be interpreted as either unconditioned stimuli or unconditioned responses. However, this interpretation cannot be applied to broader classes of sensorimotor learning, and unconditioned stimuli or unconditioned responses can be equally well interpreted as sensory errors or motor errors, which can be reduced by learned conditioned responses.

Ito and others experimentally demonstrated LTD (Ito et al., 1982; Ito and Kano, 1982; Hirano 1990; Linden and Connor, 1995; Ito, 2001). Later, long-term potentiation (LTP) was discovered when parallel-fibers were activated in the absence of climbing-fiber inputs (Hirano, 1990; 2013). Accordingly, parallel-fiber-to-Purkinje-cell synapses possess bidirectional synaptic plasticity LTD-LTP controlled by the presence and absence of climbing-fiber inputs, respectively. Earlier there was some controversy as to whether LTD-LTP is essential for several types of motor learning in the cerebellum, but there is now a consensus that it is important and necessary (Welsh et al., 2005; Schonewille et al., 2011; Anzai and Nagao, 2014; Ito et al., 2014; Yamaguchi et al., 2016; Inoshita and Hirano 2018; Kakegawa et al., 2018). If LTD-LTP is granted, it is contradictory to assign a teaching signal role to climbing-fiber inputs. Purkinje-cell simple-spike outputs would learn a sign-inverted waveform of the "teaching signal" that contradicts the "teaching signal."

In the simplest interpretation, bidirectional LTD-LTP implements the Widrow-Hoff rule for ADALINE (Widrow et al. 1976; Fujita, 1982; Kawato et al., 1987; Kawato and Gomi 1992) to minimize a squared error of regression in the steepest descent direction. The following is a mathematical demonstration that LTD-LTP can implement a simple supervised learning rule if the climbing-fiber firing modulation from its spontaneous level represents an error signal. Let us assume that Purkinje-cell simple-spike output $y$ is a linear weighted summation of parallel-fiber input $x_i$ by the parallel-fiber synaptic weight $w_i$,

$$y = \sum_i w_i x_i$$

Purkinje-cell simple-spike output $y$ learns to approximate a time-varying target value, $\hat{y}$, after sign inversion due to inhibitory action by Purkinje cells.

$$-y \to \hat{y}$$

If learning occurs in the steepest-descent direction of the squared error $E = (\hat{y} - (-y))^2 = (\hat{y} + y)^2$, the steepest-descent-direction of change in $w_i$ is as follows,

$$-\frac{dE}{dw_i} = -\frac{d(\hat{y}+y)^2}{dw_i} = -2(\hat{y}+y)x_i$$

Let $CF$ denote an instantaneous firing frequency of climbing-fiber input. $\overline{CF}$ denotes its spontaneous firing level. The basic assumption implied in Ito (1970) was that if the difference $CF - \overline{CF}$ encodes the error $\hat{y}+y$, then LTD and LTP realize steepest-descent learning with a positive learning coefficient $\varepsilon$, as follows.

$$\delta w_i = -2\varepsilon(\hat{y}+y)x_i = -2\varepsilon(CF - \overline{CF})x_i$$

This equation is in agreement with LTD (negative $\delta w_i$, and $w_i$ decreases) when $CF - \overline{CF}$ is positive (when climbing-fiber input is activated), and LTP (positive $\delta w_i$, and $w_i$ increases) when $CF - \overline{CF}$ is negative (when climbing-fiber input is silent). Considering that the baseline or spontaneous climbing-fiber firing rate is low, around 1 Hz, the negative modulation below spontaneous firing rate often lead to a fall to zero. If $CF - \overline{CF}$ can be negative then $\delta w_i$ is positive, when $CF$ is smaller than $\overline{CF}$. In other words, LTP is triggered when climbing-fiber input is silent, which is less active than spontaneous firing. This means that the negative error signal would be an inhibition of spontaneous climbing-fiber inputs (Ohmae and Medina 2015). This is the basic assumption of the unified LTP-LTD model (Kawato and Gomi, 1992).

Disinhibition of inferior olive nucleus neurons by Purkinje cells, via inhibition of the deep cerebellar nucleus, can implement the second term $y$ of the error computation $\hat{y}+y$, at least partially (Nicholson and Freeman, 2003; Rasmussen et al., 2008; Chaumont et al., 2013; Ohmae and Medina, 2015). In order for this interpretation of LTD-LTP as the error minimizing synaptic-plasticity algorithm, the neural circuit that is downstream to deep cerebellar nucleus should have a proper sign and an aligned coordinate frame so that a change in $-y$ reduces the error $\hat{y}-(-y)$. These sign and coordinate requirements are fundamental to Ito's model, and were experimentally demonstrated at the detailed neural circuit level for oculomotor adaptation, including vestibulo-ocular reflex, optokinetic responses, and ocular following responses (Simpson and Alley, 1974; Ito, 1974, 2006; Gomi and Kawato, 1992: Kawano, 1999: Kawato, 1999, Yamamoto et al., 2002). For a broader class of sensorimotor learning by the cerebellum, examination of these theoretical requirements at the neural circuit level is still technically demanding, mainly because we do not fully understand the essential low dimensions at which learning occurs. We will discuss these error and dimension issues in the Section entitled, "Toward a new computational theory of the cerebellum."

### *Coincidence detection between parallel-fiber and climbing-fiber inputs and a temporal window of synaptic plasticity*

Doya (1999) postulated that the three learning algorithms proposed in computational neuroscience, unsupervised statistical learning, reinforcement learning, and supervised learning, are the main functions of the cerebral cortex, the basal ganglia, and the cerebellum, respectively (see also Raymond and Medina (2018) for cerebellar supervised learning). In the Section entitled, "Toward a new computational theory of the cerebellum", we argue against this classification and propose that all three brain regions are for reinforcement learning. However, here we agree with Doya (1999), in the sense that a local neural circuit around Purkinje cells is most suited for supervised learning, based on a biophysical model of LTD for parallel-fiber-to-Purkinje-cell synapses (Figures 1 A and B). Large calcium increases in dendritic spines induce long-term decreases of synaptic efficacy, LTD, in the cerebellum (Figure 1C), while it induces long-term increases of synaptic efficacy, LTP, in the cerebral cortex (Figure 1D). In contrast, small calcium increases induce LTD in the cerebral cortex, while by themselves, they do not induce LTP in the cerebellum (Tanaka et al., 2007). Consequently, other factors such as nitric oxide are necessary for LTP induction in the cerebellum (Ogasawara et al., 2007).

Figure 1A depicts schematically the early phase of LTD of parallel-fiber-Purkinje-cell synapses up to a large calcium increase (Kuroda et al., 2001; Ito, 2002; Kotaleski et al., 2002; Doi et al., 2005). Parallel-fiber inputs depolarize dendritic spines via *α-amino-3*-hydroxy-5-methyl-4-isoxazolepropionic acid receptors (AMPARs). In parallel to this action, glutamate released from parallel fibers binds to metabotropic glutamate receptor type 1 (mGluR1) inducing a slow increase of inositol 1,4,5-triphosphate ($IP_3$) with 100 millisecond-order time to peak, via G-proteins and phospholipase $C\beta$. On the other hand, climbing-fiber inputs, which lag about 100 milliseconds behind parallel-fiber inputs, induce large depolarizations in dendrites via multiple strong excitatory synapses, thereby opening voltage-dependent calcium channels on the spine and inducing calcium influx (Ito, 2002; Doi et al., 2005) (Figure 1A). Because the latter electrical event is much faster than the former, $IP_3$ and $Ca_{2+}$ concentrations increase simultaneously in the spine. This triggers a regenerative $Ca_{2+}$ increase by $IP_3$-induced $Ca_{2+}$ release (IICR) through $IP_3$–bound $IP_3$ receptors ($IP_3Rs$). $IP_3Rs$ are $IP_3$-gated $Ca_{2+}$ channels on the endoplasmic reticulum (ER), which is the intracellular $Ca_{2+}$ store. IICR results in a supralinear $Ca_{2+}$ surge with several micro-molar peaks (Wang et al., 2000; Doi et al., 2005) (Figure 1B).

The $Ca_{2+}$ surge induces subsequent reactions including activation of a mitogen-activated protein kinase (MAPK)-positive feedback loop, which phosphorylates and then internalizes AMPARs, and induces and consolidates LTD (Kuroda et al., 2001). Tanaka et al. (2007) demonstrated that a MAPK-positive feedback loop operates as a leaky integrator of $Ca_{2+}$ concentration to determine the threshold of LTD, thus determining the temporal window of LTD. $IP_3Rs$ and $IP_3$-dependent IICR act as coincidence detectors of the preceding parallel-fiber and delayed climbing-fiber inputs, as detailed in the following. Because the opening probability of $IP_3Rs$ increases with $IP_3$ concentration, the $Ca_{2+}$ threshold for IICR $Ca_{2+}$ surge decreases within a few hundreds of milliseconds (blue curve in Figure 1B) after multiple parallel-fiber inputs as $IP_3$ concentration slowly increases via the mGluRs pathway, typically achieving its minimum after 100 milliseconds (Figure 1B). $Ca_{2+}$ increase by climbing-fiber inputs around this minimum-threshold time is sufficiently large to cross the threshold (thick black curve in Figure 1B) so that IICR $Ca_{2+}$ surge is triggered and drives LTD as explained above. If climbing-fiber inputs are either too early (gray curve in Figure 1B) or too late (dotted curve in Figure 1B) with respect to this minimum-threshold time, then the $Ca_{2+}$ increase is below the threshold of the IICR $Ca_{2+}$ surge and fails to induce LTD (Figure 1B). In summary, the temporal window of LTD is determined by a slow increase in $IP_3$ concentration and the resultant dynamic threshold change determined by $IP_3Rs$ kinetics, and is on the order of 100 milliseconds. Simplistically speaking, the molecular cascade consisting of glutamate, mGluR1, G-proteins, phospholipase $C\beta$, $IP_3$, IICR of $IP_3Rs$ generates the delay such that the climbing fiber signal coming later can fall in the temporal window of LTD.

In addition, the intensity of climbing-fiber input can matter. Complex spikes measured in extracellular recording are not homogeneous and consist of 1-6 consecutive spikes of climbing-fibers (called spikelets; Armstrong and Rawson, 1979; Maruta et al.,2007; Mathy et al., 2009). The number of spikelets is essential to induce motor learning (Rasmussen et al., 2013; Yang and Lisberger, 2014). As the number of spikelets increases, it may reach the threshold of $Ca_{2+}$ spikes, which could not be reached by shorter spikelets, and may extend the temporal window of $Ca_{2+}$ surge and LTD. On the other hand, in saccadic adaptation, the timing of climbing-fiber input on the order of 100 milliseconds is suggested to be more important than the number of spikelets (Herzfeld et al., 2018). The ability of spikelet to extend the temporal window may vary, depending on the context and/or the area in the cerebellum.

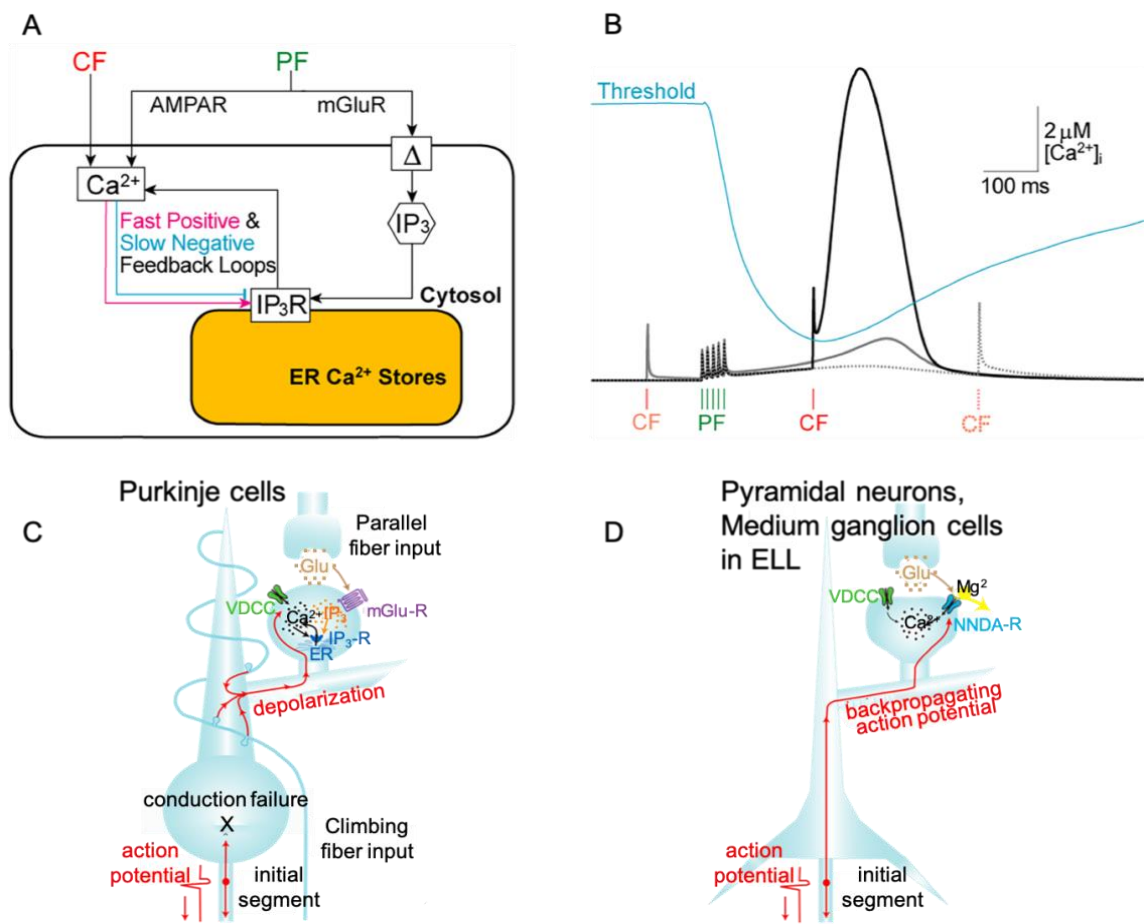

**Figure 1. Different synaptic plasticity mechanisms of Purkinje versus pyramidal cells. A.** Schematic diagram of an LTD biophysical model (Doi et al. 2005). **B.** Simulation reproduces the temporal window of Ca2+ surge that triggers LTD. **C.** An LTD biophysical model overlaid on morphology, key molecules, and physiology of Purkinje cells. **D.** Synaptic plasticity and coincidence detection mechanisms in pyramidal neurons in the cerebral cortex and medium ganglion cells in electrical lobes (ELL) in comparison with Purkinje cells in the same format as shown in C.

*Figure **A** and **B** were published in J Neurosci 25, Doi T, Kuroda S, Michikawa T, Kawato M, Inositol 1,4,5-trisphosphate-dependent Ca2+threshold dynamics detect spike timing in cerebellar Purkinje cells, 950-961, Copyright The journal of Neuroscience (2005).

*Figure **C** and **D** were published in Curr Opin Neurobiol 21(5), Kawato M, Kuroda S, Schweighofer N, Cerebellar supervised learning revisited: biophysical modeling and degrees-of-freedom control, 791-800, Copyright Elsevier (2011).

*Contrasting synaptic plasticity mechanisms between Purkinje cells and pyramidal neurons*

Figures 1 C and D contrast different synaptic plasticity mechanisms of Purkinje cells of the cerebellum and pyramidal neurons of the cerebral cortex, while illustrating different morphological, physiological, and molecular mechanisms, with emphasis on coincidence

detection. The large $Ca_{2+}$ surge in Purkinje cells is induced by IICR from IP$_3$Rs and is mainly triggered by calcium influx caused by climbing-fiber inputs. Note that action potentials in Purkinje cells do not backpropagate because of excessive electrical load by extensive dendrite branching (Vetter et al., 2001) and the low density of sodium channels on dendrites (Llinas and Sugimori, 1980; Stuart and Häusser, 1994). Consequently, whether Purkinje cells are excited is irrelevant to LTD, while the large depolarization and subsequent $Ca_{2+}$ increase induced by climbing-fiber inputs are decisive for LTD occurrence. Thus, supervised learning guided by error signals is suggested for the cerebellum. In contrast, in cerebral pyramidal neurons (Figure 1D), NMDA receptors (*N*-methyl *D*-aspartate receptor, NMDAR) are coincidence detectors of glutamate released from presynaptic terminals and the backpropagating action potential from the axon initial segment (Linden, 1999; Caporale and Dan, 2008). Glutamate, released from presynaptic terminals, binds to NMDARs, and backpropagating action potentials increase the postsynaptic voltage and consequently release a $Mg_{2+}$-block of glutamate-bound NMDARs, resulting in full activation of NMDARs (Linden, 1999; Caporale and Dan, 2008; Urakubo et al., 2009). This leads to a large $Ca_{2+}$ influx via NMDARs and induces subsequent reactions to consolidate LTP. Because the release of NMDARs from the $Mg_{2+}$-block by backpropagating action potentials is the decisive event that leads to large calcium influx, Hebbian and unsupervised statistical learning are suggested for the cerebrum (Figure 1D). For parallel-fiber-Purkinje-cell synapses of adult cerebellum, there is no evidence that NMDARs in Purkinje cells participate significantly in LTD (Perkel et al., 1990; Renzi et al., 2007; Kono et al., 2019). In striatal medium spiny neurons, while $Ca_{2+}$ influx depends on NMDAR activation by backpropagating action potentials, as in the cerebrum, synaptic plasticity also depends on activation of dopamine receptors (Wickens et al., 1996; Shen et al., 2008). In D1 receptor-expressing neurons, activation of the positive feedback loop, composed of PKA, PP2A and DARPP-32, serves as the coincidence detector of $Ca_{2+}$ influx and dopamine input (Lindskog et al., 2006; Fernandez et al., 2006; Nakano et al., 2010); thus, the reinforcement learning rule is supported. Because D1 receptors and DARPP-32 are expressed in prefrontal cortex, but at lower levels, the positive feedback loop probably does not possess bistability between the basal state and the excited state; thus, it cannot implement reinforcement learning rules, as in the basal ganglia. This is because any reinforcement learning rule necessitates memory functions provided by some kinds of bistability. In summary, Purkinje cell LTD is unique, supervised, and not Hebbian, while all those in the cerebral cortex, hippocampus, and basal ganglia are basically Hebbian, meaning that action potentials of the post-synaptic neurons are essential for synaptic plasticity.

    According to the different mechanisms of coincidence detection in synaptic plasticity for the cerebellar cortex and the cerebral cortex, temporal windows of synaptic plasticity are also qualitatively and quantitatively different. The synaptic eligibility trace is a record of a synapse's recent activities to mark it as being eligible to change and to become distinct from other synapses (Barto et al., 1983). The synaptic eligibility trace for Purkinje cells is slowly rising IP$_3$ (Okubo et al., 2004; Doi et al. 2005); thus, the temporal window of LTD with respect to parallel-fiber input is ~100 milliseconds in the cerebellum. In contrast, NMDAR dynamics are the synaptic eligibility trace for pyramidal neurons, dynamics that determine a temporal window of ~10 milliseconds (Markram et al., 1997). Furthermore, when synaptic input precedes postsynaptic firing, LTP is induced, whereas if the synaptic input lags behind the firing, LTD is induced, which is spike-timing-dependent plasticity (STDP) (Bi and Poo, 2001). While late phases of LTD of the cerebellum and LTP of the cerebrum seem to share common signal-transduction mechanisms, the early phases are distinctly different, as explained above, and partly explain differences in learning algorithms at a micro level, as in dendritic spines.

### *Meta-plasticity of the temporal window of Purkinje-cell LTD*
In the framework of Ito's model, simple spikes of Purkinje cells elicited by a combination of parallel-fiber inputs and spontaneous activities represent motor commands, and complex spikes of Purkinje cells elicited by climbing-fiber inputs represent error signals caused by executed movements. Thus, complex spikes should be delayed with respect to simple spikes, because of feedback delays. This delay includes latency for conduction of command signals within a neural circuit downstream to the cerebellum, muscle stretches, movement execution, sensory organ activation, afferent signal conduction, and finally activation of inferior olive neurons. Yamamoto and colleagues (2002) assumed that this delay was compensated by the temporal window of LTD and conducted efficient learning simulations. That is, the temporal correspondence between the responsible motor command (parallel-fiber inputs and simple spikes) and the resulting motor error (climbing-fiber inputs and complex spikes) is guaranteed in LTD, if the motor-sensory time delay is matched to the temporal window of LTD (Ogasawara et al., 2008). The motor-sensory time delay should be different for different cerebellar regions controlling different movements. For example, the delay is expected to be small, around 30 milliseconds, for ventral paraflocculus controlling ocular-following responses. The delay is expected to be large and longer than 100 milliseconds for a region in a hemisphere responsible for visually guided arm-reaching. The size (width and center) of the LTD temporal window for different parts of the cerebellum should be matched to the motor-sensory time delays, if the objective of the LTD temporal window is to cancel the delay. Suvrathan and colleagues (2016, 2018) found experimental data supporting this theoretical requirement. In the biophysical model of LTD (Doi et al., 2005), the optimal time delay between parallel-fiber inputs and climbing-fiber inputs for effective LTD, the time for $IP_3$ concentration to reach its maximum after parallel-fiber inputs arrive, is determined by a biochemical reaction delay in the mGluR pathway. This biochemical delay depends on concentrations of the molecules involved, three-dimensional arrangements of anchor proteins, diffusion constants of molecules, and geometrical characteristics of cytosols within dendritic spines. It would be exciting research to theoretically and experimentally investigate possible chemical mechanisms for meta-plasticity to change some of the above factors so that the biochemical delay is matched to the neural delay in a region- and movement-specific manner in the cerebellum.

## Codon theory

Marr and Albus proposed "Codon Theory" for representations and computations by granule cells, while Ito did not address this subject (Table 5). Because Brindley (1964, 1969), a mentor of Marr, already discussed cerebellar-supervised learning, codon theory is one of the most innovative elements of Marr's model. The number of cerebellar granule cells is ~50 billion, exceeding the sum of all other neurons in the brain. Each granule cell possesses 4 to 5 synapses on small dendrites and receives synaptic inputs from mossy-fibers. The number of granule cells is 200 times the number of mossy-fibers, ~250 million (Figure 2B). This corresponds to "expansion coding" in modern terminology. In modern computational terminology, codon theory also postulates "sparse coding" such that only a small number of granule cells are activated when specific combinations of their mossy-fiber inputs are activated. By expanding the dimension of input representation (expansion) and reduction of the number of simultaneously activated granule cells (sparseness), interference of associative memories for different contexts is minimized; thus, efficient learning can be expected.

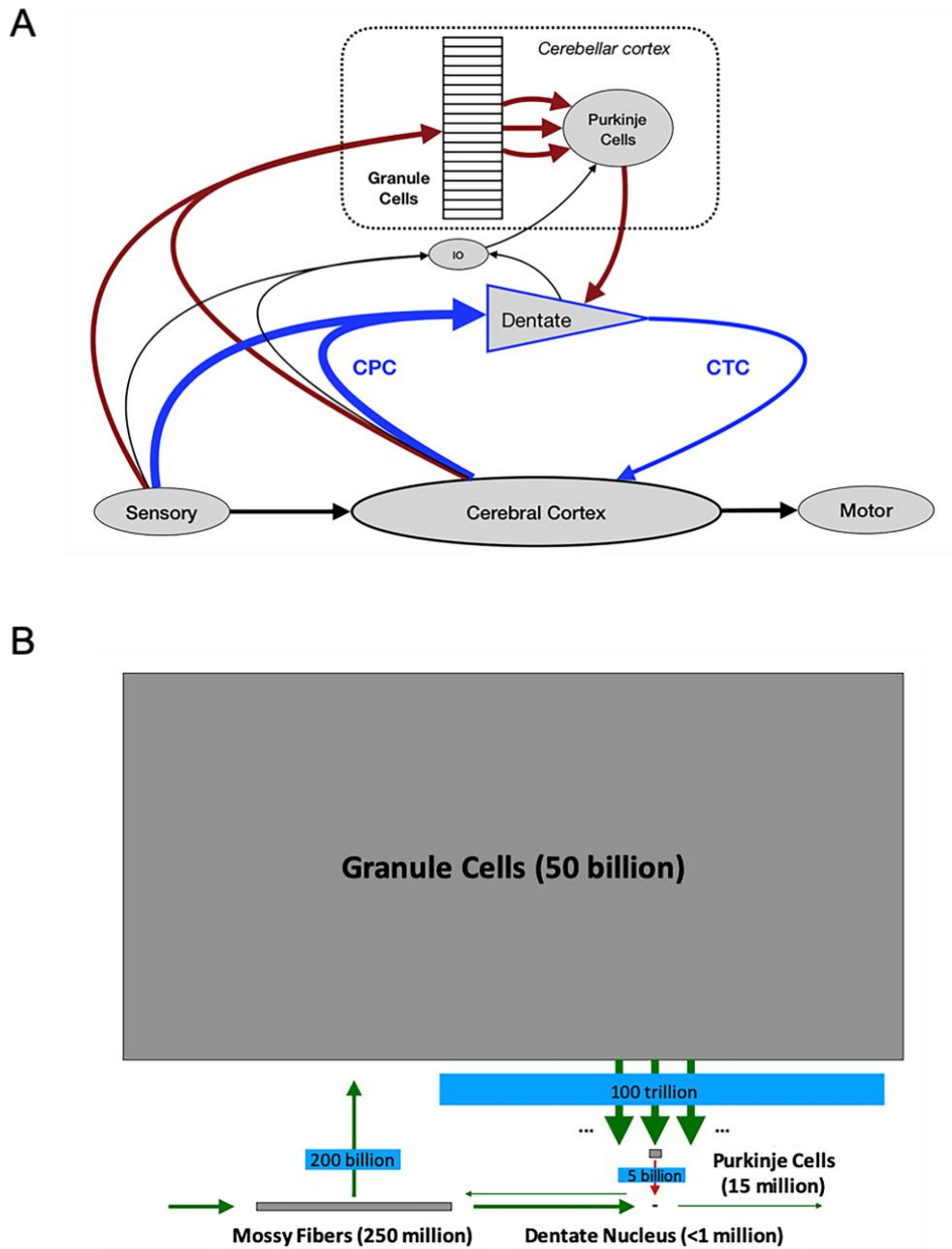

**Figure 2. A. Circuit diagram of the cerebellum with the dentate nucleus, emphasizing its parallel path characteristics to the cerebral cortex in sensorimotor control (taken from Sanger et al., 2020).** In **A**, IO: inferior olive nucleus, CPC; cortico-ponto-cerebellar pathway, CTC; cerebello-thalamo-cortical pathway. **B. Relative numbers of cells and synapses in mossy-fibers, granule cells, Purkinje cells, and dentate-nucleus circuits (taken from Sanger et al., 2020).** In **B**, the area of each gray rectangle is proportional to the approximate cell count for the corresponding cell type. The area of each blue inset illustrates the relative number of output synapses from that type of neurons.
*Figure 2 was published in Journal of Physiology, 598(5), Sanger TD, Yamashita O, Kawato M, Expansion coding and computation in the cerebellum: 50 years after the Marr-Albus codon theory, 913-928, Copyright John Wiley and Sons (2020).

Codon theory predicts nonlinear input-output transformation at granule cells, low firing rates of granule cells, activation of only a small portion of all granule cells, and multiple modalities of mossy-fiber inputs to individual granule cells, at least for certain regions of the cerebellum and for certain cerebellar functions. In the past two decades, large technical advances of intracellular and patch clamp recordings, as well as $Ca_{2+}$ imaging from granule cells have been made. Contradictory data for the above four predictions has been obtained and some have even argued that codon theory has been disproven. Sanger et al. (2020) recently reviewed this evidence and pointed out that seemingly contradictory data obtained under specific preparations and conditions cannot disprove the core of codon theory. Because of this recent review, here we only very briefly discuss the above four points: nonlinear transformation, low firing rates, sparse coding, and multiple input modality.

Under some experimental conditions, granule cells can fire at high rates when activated by only a single mossy-fiber, and the input-output firing-rate relationship is linear (Jörntell and Ekerot, 2006; Rancz et al., 2007; Arenz et al., 2008; van Beugen et al., 2013). Kawano and colleagues showed that instantaneous firing rates of both mossy-fibers and parallel fibers are related to eye movement dynamics and are similar during ocular following responses in ventral paraflocculus (Kawano and Shidara, 1993; Kawano et al., 1996a, 1996b). Head rotation velocities were predicted from mossy-fiber firing frequencies to granule cells, suggesting that instantaneous firing rates encode information in granule cells (Arenz et al., 2008). Recent $Ca_{2+}$ imaging studies indicated that a large proportion of granule cells are excited simultaneously, which appears to contradict sparse coding (Spanne and Jörntell, 2015; Badura and De Zeeuw, 2017; Giovannucci et al., 2017; Knogler et al., 2017; Wagner et al., 2017; Gilmer and Person, 2018). However, this point critically depends on temporal resolution, or on a minimal meaningful time bin of granule cell encoding. In some examples of eye movements, temporal resolution of simple-spike firing is around 1 millisecond (Shidara et al., 1993; Gomi et al., 1998; Payne et al., 2019). If a sampling frequency of $Ca_{2+}$ imaging is 30 Hz and 50% of granule cells are co-activated in the imaging resolution, minimally only 1.5% of granule cells are simultaneously activated in a 1-millisecond time bin. For the vestibular cerebellum and regions that directly receive tactile inputs, under decerebrated and/or anesthetized conditions, granule cells receive synaptic inputs from a single modality (Jörntell and Ekerot, 2006; Bengtsson and Jörntell, 2009). However, Ishikawa et al. (2015) found that mossy-fiber inputs of somatosensory, auditory, and visual modalities from the cerebral cortex converge on individual granule cells (Chabrol et al., 2015; Shimuta et al., 2019). In this case, nonlinear input-output relationships were observed for combined stimuli. Anatomically, one mossy-fiber carrying proprioception from the external cuneate, and another mossy-fiber carrying other modalities from pontine nuclei converge to a single granule cell (Huang et al., 2013).

In the Section, Comparisons of models by Marr, Ito, and Albus, we discussed pattern recognition versus control functions of the cerebellum. An all-or-nothing (firing or not) representation is best suited for pattern recognition, and codon theory is based on these assumptions. Even for instantaneous firing-rate coding and control (regression), codon theory might be relevant to enhancement of approximation precision in regression, as follows. Neuronal origins of mossy-fibers, granule cells, and Purkinje cells form a three-layer feedforward neural network. Funahashi (1989) and Cybenko (1989) mathematically proved that such three-layer neural networks can learn to approximate any arbitrary, smooth, nonlinear functions if the intermediate layer has a sufficiently large number of neurons, and synaptic weights are optimally chosen. As will be discussed in the Section, Internal models, function approximation capability is essential for learning acquisition of forward models or inverse models of nonlinear, controlled objects within the three-layer network. Because mossy-fiber-granule-cell synapses possess

Hebbian plasticity (Schweighofer et al., 2001; Sgritta et al., 2017), backpropagation learning algorithms used for deep neural networks cannot be applied. Consequently, in order to increase function approximation capability of the three-layer network, a huge number of granule cells and a rich repertoire of nonlinear basis functions are essential there (Sanger et al., 2020). In this sense, the spirit of codon theory, which was proposed in pattern recognition domains, is still relevant even for cerebellar functions and regions related to control, regression, and instantaneous firing-rate coding. This is because the rich repertoire of nonlinear transformations, provided by a vast number of granule cells, is essential for approximation capabilities by the cerebellum, and it is a core assumption of codon theory.

Higher expansion ratios are beneficial for more elaborate cerebellar functions dealing with strong nonlinearity; thus, the relevance of codon theory is higher for larger expansion ratios. Recent structural magnetic resonance imaging studies on gray matter and input fibers of human cerebellum suggested that expansion ratios of granule cells from mossy fibers are different between the vermis and hemispheric regions. The gray-matter volume of the hemisphere was 11.4x of that of the vermis, while the cross-sectional area of the cerebrocerebellar peduncle was estimated at less than 2.8x the spinocerebellar tract (Keser et al., 2015). On the assumption that diameters of mossy-fibers of these two regions of the cerebellum do not differ, we estimate that the expansion ratio of the hemisphere is 4x larger than that of the vermis. Large expansion ratios might be more marked for phylogenetically newer parts of the cerebellum and higher cognitive functions. In contrast, seemingly contradictory data against codon theory were mostly obtained in older parts of the cerebellum under anesthetized or decerebrate preparations. We postulate that codon theory is more relevant to newer functions of the cerebellum.

## Internal models

### Cerebellar internal models

Internal models in neuroscience are the neural networks that simulate input-output relationships of some processes, such as controlled objects in movement execution. "Internal" implies that the neural network is within the brain or the cerebellum. "Model" implies that the neural network simulates a target dynamical process. When humans move their bodies quickly, the necessary computation is too difficult to be solved by simple feedback controllers alone, and internal models are necessary (Gomi and Kawato, 1996). Here, we include both forward and inverse (see below) models as possibilities. There is always feedback control, but except very basic feedback control, model-predictive controllers or precomputed ballistic movements include some forms of internal models. When humans grow their bodies and manipulate different objects, their dynamics change drastically; thus, internal models cannot be genetically preprogrammed or fixed, and must be acquired through learning. Ito (1970) proposed that internal models are acquired by learning in the cerebellum (Table 1; Figures 3A and B). Neither Marr nor Albus mention internal models around 1970, but Albus (1975) later proposed an artificial neural network model called CMAC (cerebellar model articulation controller), and one of its possible applications was to learn inverse dynamics models of robots.

Internal models are classified as either forward or inverse. Forward models possess the same input-output direction as controlled objects and simulate their dynamics. For example, controlled objects such as eyeballs or arms receive motor commands and generate movement trajectories, that can be represented as sensory feedback about the executed trajectories. Forward models receive copies of motor commands (corollary discharge or efference copy) and predict movement trajectories or sensory feedback (sensory consequence). In control engineering and robotics, forward models have been and still are just called "internal models." Jordan and Rumelhart (1992) coined this term to discriminate them from inverse models, and this terminology was soon adopted by cognitive science and neuroscience. On the other hand, inverse models simulate inverted input-output relationships of controlled objects. In a sense, inverse models provide inverse functions of modeled dynamical systems. Inverse models of controlled objects can receive desired trajectories as inputs and can compute necessary motor commands to realize the desired trajectories. Because an inverse model possesses the inverse input-output relationship of a controlled object, a cascade of the inverse model and the controlled object becomes an identity map. This implies that the trajectory realized by this cascade is equal to the desired trajectory; thus, the inverse model provides an ideal feedforward controller. We note that forward models are well-posed problems (every action causes a unique result) whereas inverse models are not (the same result can occur following many actions). So, among many possibilities, the brain needs to select one for the inverse model to learn. Although the combination of an inverse model followed by a forward model is the identity, it is not unique.

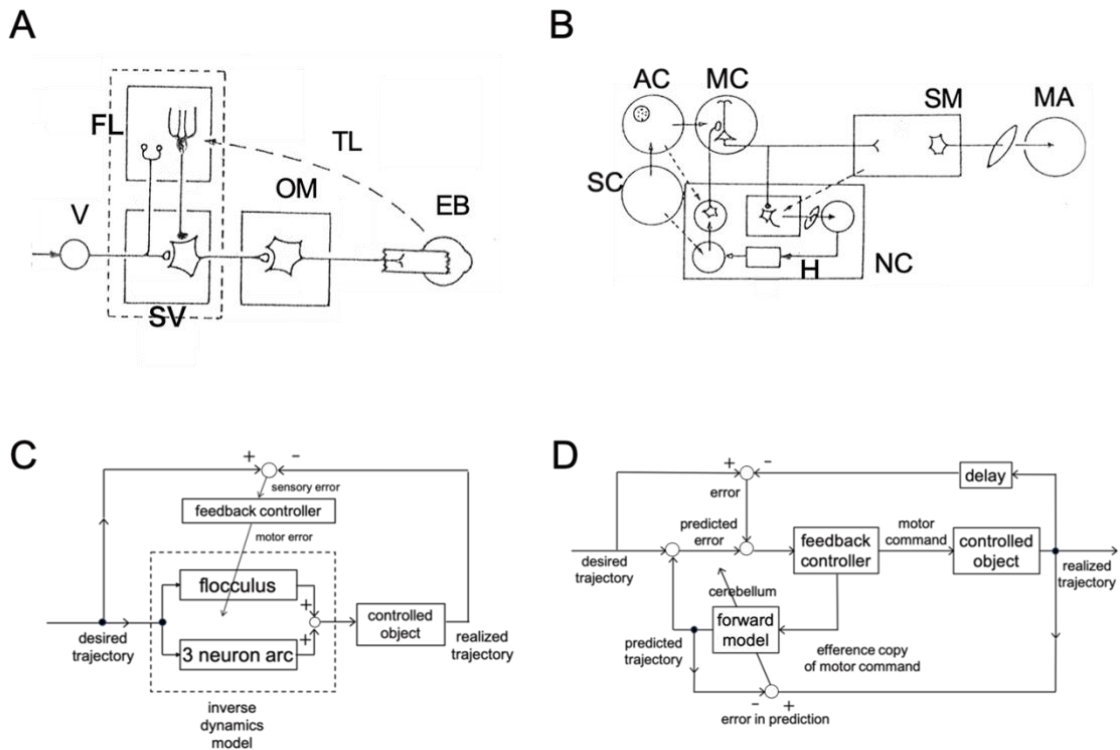

**Figure 3. Figures from Ito (1970) and interpretations thereof, based on inverse and forward internal models.** **A** and **B** are two figures from Ito (1970), and **C** and **D** are their interpretations, respectively. **A**. Figure 6C of Ito (1970) for feedforward control of the vestibulo-ocular reflex. FL, flocculus, V, sensory input from the vestibular organ, SV, rostral part of the vestibular nuclear complex, OM, oculomotor neuron, EB, eyeball, TL, teaching line for the learning in cerebellar cortex. **B**. Figure 7B of Ito (1970) for the control system of voluntary movements. AC, cerebral association area, MC, cerebral motor area, SM, spinal motor system, SC cerebral sensory area, MA, motor activity, H, internalized feedback pathway through the external world, NC, neocerebellum in which SM, MA, H, SC, and AC are modeled in a miniaturized form. **C**. Inverse dynamics model interpretation of **A.** for adaptation of the vestibulo-ocular reflex. FL and V-SV-OM in **A** correspond to flocculus and 3 neuron arc in **C**, respectively. Dashed rectangles in **A** and **C** correspond with each other. The flocculus and the 3-neuron arc function in parallel to acquire an inverse dynamics model of the controlled object. A feedback controller for optokinetic

responses in the accessory optic system transforms a retinal slip into the motor error, which is used for adaptive changes of the inverse dynamics model in the flocculus. **D**. Forward internal model interpretation of **B**. The neocerebellum acquires a forward model of a controlled object, and uses it for internal feedback control. NC in **B** corresponds to forward model in **D**. MC in **B** corresponds to feedback controller in **D**. H in **B** corresponds to upper feedback loop in **D**.

Figures 3C and D show our interpretations of Figures 3A and B proposed by Ito, while utilizing modern terminology and computational backgrounds of inverse and forward models, respectively. In Figure 3C, the 3-neuron vestibulo-ocular reflex arc and the flocculus form a parallel system and the combined system provides an inverse dynamics model of a controlled object, a downstream of the oculomotor system from the vestibular nucleus. Note that the 3-neuron arc is fixed and the flocculus is modifiable; thus, the combined system is modifiable and can learn the inverse dynamics model. The sensory error between the desired trajectory (sign inverted head rotation) and the realized trajectory (eye rotation) is given as a retinal slips and it is transformed into motor coordinates by the control network of optokinetic responses, and used as the motor command error to guide learning of a part of the inverse dynamics model in the flocculus (Gomi and Kawato, 1992; Tabata et al., 2002). Figure 3B can be also interpreted to mean that a newer part of the cerebellum provides an inverse dynamics model of a controlled object. In this interpretation, a sensorimotor cortex of the cerebrum provides a feedback motor command produced through long-loop sensory feedback through an external world, and this feedback motor command is utilized to train the inverse dynamics model in the cerebellum, that is, it functions as an ideal feedforward controller. In another interpretation based on the forward model (Figure 3D), part of the cerebellar hemisphere provides a forward model of the controlled object. It computes a predicted trajectory while receiving the efference copy of the motor command. The difference of the realized trajectory and the predicted trajectory provides a sensory prediction error that can be used as an error signal to train the forward model. This forward model can be used as an essential element for internal feedback control, bypassing the long-loop through the external world, reducing feedback delays and increasing control performance. Figure 3D illustrates this latter interpretation of Figure 3B based on later computational models of internal feedback control with forward models (Kawato et al., 1987; Kawato and Gomi, 1992; Miall et al., 1993). A linear quadratic regulator (Bemporad et al., 2002; Li and Todorov, 2004) and an optimal feedback controller proposed by Todorov and Jordan (2002) can be regarded as more sophisticated feedback control algorithms, which also utilize forward models.

As discussed in the Section, Codon theory, the 3-layer feedforward neural circuit comprising neuronal origins of mossy-fibers, granule cells, and Purkinje cells, is expected to closely approximate arbitrarily complex nonlinear functions with a vast number of granule cells, if an appropriate error signal is provided and optimal synaptic weights are learned. Thus, the cerebellum basically possesses learning capability to acquire internal models if climbing-fiber inputs provide appropriate error signals. Kawato and colleagues (1987) proposed that forward and inverse models are acquired distinctively in different parts of the cerebellum. Learning of forward models is computationally straightforward, since the error signal can be computed as the difference between the realized trajectory (or sensory feedback) and the predicted trajectory (or predicted sensory feedback) (Figure 3D) (see also discussion of computing $\hat{y} + y$ by Purkinje-cell disinhibition in the Section, Synaptic plasticity in achieving supervised learning). In contrast, learning of inverse models is fundamentally difficult, because we cannot assume the existence of a teaching signal $\hat{y}$, i.e., ideal motor commands in the brain. This is because if such an ideal motor command already exists, it can be used as the appropriate motor command in movement executions, and inverse models or learning them is unnecessary. We postulated a "feedback-error-learning" scheme in which the inverse dynamics model receives the error signal in a motor command coordinate from the output of the feedback controller, which performs more crude and poorer sensorimotor transformation than the inverse model can do, but still provides the necessary motor error (Kawato et al., 1987). The feedback controller receives a sensory error as its input, and the output from the feedback controller can be regarded as the motor error necessary for training the inverse model (Figure 3C). Accordingly, climbing-fiber inputs for some regions were postulated to represent outputs of feedback controllers (Kawato and Gomi, 1992; Gomi and Kawato, 1992; Bhushan and Shadmehr, 1999; Tabata et al., 2002; Ramnani, 2006; Ito 2008). Inoue and colleagues (2016, 2018) showed that motor commands of cerebral motor cortex in non-human primates can drive motor adaptation and can be potential sources of error signals for the cerebellum, in accordance with the feedback-error-learning scheme.

*Experimental support for forward and inverse models*
Experimentally, it is not easy to address whether the cerebellum is predicting the sensory consequence of the movement (forward model) or computing the necessary feedforward motor command (inverse model). Fundamental mathematical difficulties in asking this question were demonstrated by Mehta and Schaal (2002). It has been reported that the activity of cerebellar Purkinje cells and cerebellar nuclei encode kinematics (position and/or velocity) of movement, suggesting forward models in the cerebellum (Laurence et al., 2013; Brooks and Cullen, 2013; Brooks et al., 2015; Herzfeld et al., 2015, 2018; Streng et al, 2018). However, since signals of position or velocity can be a part of the motor command, the possibility of signaling by inverse models cannot be excluded. Similarly, some experimental data demonstrated that Purkinje cell activities directly drive movements and support inverse models rather than forward models (Heiney et al., 2014; Bhanpuri et al., 2014; Hoogland et al., 2015; Lee et al., 2015; El-Shamayleh et al., 2017; Chabrol et al., 2019; Payne et al., 2019). However, it is also possible that stimulation interferes with normal prediction of sensory consequences of movement and induces artifactual movements by sensori-motor transformation in the downstream of the cerebellum. To solve the problem of discriminating forward from inverse models, carefully designed paradigms are required to dissociate movement consequences and motor commands. Pasalar et al (2006) and Yamamoto et al (2007) are representative examples of such studies in non-human primates, but surprisingly, they drew opposing conclusions. The former supported the forward model, while the latter supported the inverse model. This discrepancy can be explained by assuming that slow movement in the former study does not require feedforward control by an inverse model, whereas feedforward control was essential for much faster movements in the latter study. Although other research also supported cerebellar forward models (Wolpert et al., 1995; Blakemore et al., 2001; Kawato et al., 2003; Bastian 2006; Tseng et al., 2007; Izawa et al., 2012; Sultan et al., 2012;), experimental data for cerebellar inverse models are more straightforward in their interpretations. In particular, various studies on ocular following responses in the ventral paraflocculus (see the Section, Synaptic plasticity in achieving supervised learning) have demonstrated that Purkinje cell input and output correspond to sensory and motor coordinates, respectively, and that the inverse model is computed in the cerebellum (Shidara et al., 1993; Kawano et al., 1996a, 1996b; Gomi et al., 1998; Kobayashi et al., 1998; Kawano, 1999; Kawato, 1999; Yamamoto et al., 2002; Takemura et al., 2017).

*Electrosensory lobe and forward models*
One of the important functions of forward models is central cancelation of sensory consequences of autogenous movement. The electrosensory lobe of weakly electric fish is one of many cerebellum-like structures, and illustrates beautifully this proposed function. Medium ganglion cells of the electrosensory lobe estimate electrosensory signals induced by external events by subtracting electrosensory signals caused by the fish's own electric organ discharge and movements (Bell and Russell, 1978; Han et al., 2000; Requarth and Sawtell, 2011; Kennedy et al., 2014; Sawtell 2017). This can be regarded as a special case of estimating sensory inputs. Forward models can predict sensory inputs caused by an animal's own movement from the efference copy of motor commands. Because of this finding, some argue that the cerebellum should also cancel sensory inputs caused by autogenous movements utilizing forward models, i.e., central cancelation of sensory consequences of autogenous movements by forward models. However, this expectation is not fully supported when we consider different synaptic plasticity mechanisms between medium ganglion cells and Purkinje cells and different representations of climbing-fiber inputs to different types of neurons. In Figures 1C and D, we contrasted different synaptic plasticity mechanisms of Purkinje cells and pyramidal neurons of the cerebrum. Synaptic plasticity rules of medium ganglion cells are anti-Hebbian and are more similar to those of pyramidal neurons than of Purkinje cells. We note that anti-Hebbian learning requires action potential generation in post-synaptic neurons for depression of synapses, which is different from Purkinje cell LTD. As already explained, action potentials do not backpropagate in Purkinje cells, and so far there is no evidence that post-synaptic firing in Purkinje cells contributes to the synaptic plasticity of LTD. Climbing-fiber inputs to medium ganglion cells represent total electrosensory signals, and parallel-fiber inputs include motor corollary discharge signals related to the electric organ discharge, and proprioceptive signals related to movements and position of the tail, trunk, and fins. When the medium ganglion cell, and parallel-fiber synapse carrying the corollary discharge, and climbing-fiber input are all simultaneously activated, according to anti-Hebbian learning, parallel-fiber synapses undergo LTD. Consequently, learning cancels the electrosensory signal that can be predicted by the corollary discharge from the firing output of the medium ganglion cell. As explained in the Section, Synaptic plasticity in achieving supervised learning, this anti-Hebbian learning is completely different from the supervised learning rule of Purkinje cells, and the same cancelation computation is impossible for Purkinje cells. If climbing-fiber inputs to Purkinje cells encode the error between the actual sensory input and that predicted by the corollary discharge, then climbing-fiber inputs should decrease to zero while Purkinje cells continue to fire, if there is no external sensory event after learning is completed. In contrast, climbing-fiber inputs to medium ganglion cells continue to fire while medium ganglion cells become silent when there is no external sensory event after learning is completed. Accordingly, both upstream and downstream neural circuits represent markedly different pieces of information, so they should be wired entirely differently for the cerebellum and the electrosensory lobe. From computational viewpoints, Purkinje cells can perform sensori-motor transformation, whereas medium ganglion cells simply subtract climbing-fiber inputs from parallel-fiber inputs within the same dimensions without coordinate transformation. We acknowledge that some parts of the cerebellum most probably acquire forward internal models for central cancelation of sensory consequences of autogenous movement (Blakemore et al., 1998, 2001; Wolpert et al., 1998; Kawato et al., 2003; Brooks and Cullen, 2013; Brooks et al., 2015), but experimental and computational evidence from the electrosensory lobe is not directly relevant to this function of the cerebellum.

### *Cerebellar internal models for cognition*
Cerebellar internal models can simulate any dynamical processes other than controlled objects in sensorimotor tasks, but only if climbing-fibers represent appropriate error signals between the actual and predicted outputs from the dynamical processes, and possibly provide computational mechanisms for recent proposals of cerebellar cognitive functions (Ito 2008; Strick et al., 2009; Lu et al., 2012; Honda et al., 2018; Schmahmann et al., 2019; Deverett et al., 2019). Higher cognitive functions are realized in the brain, utilizing cerebral cortex, basal ganglia, and subcortical structures. If inputs and outputs of these computational modules are provided to newer parts of the cerebellum as parallel-fiber and climbing-fiber inputs, different parts of the cerebellum can learn to acquire either forward or inverse models of these cognitive processes. Imamizu and Higuchi showed that internal models of many different tools are acquired in the human cerebellum in a modular manner, and that their representations largely overlap with those for language in the cerebellum, as well as in Broca's area of the cerebral cortex (Imamizu et al., 2000, 2003; Higuchi et al., 2007, 2009). If multiple internal models are acquired in a modular and switching mode suggested by mixture-of-experts model of artificial neural networks (Jacobs et al., 1991), they are advantageous to cope with very complicated tasks such as rapidly switching manipulation of many different objects (Gomi and Kawato, 1993). A basic form of Modular Selection And Identification for Control (MOSAIC) architecture consists of multiple paired forward and inverse models, and can be regarded as a more parallelized version of the mixture of experts architecture (Wolpert and Kawato, 1998). Each forward model tries to predict local dynamics and competes with other forward models for goodness of prediction, which is measured as a responsibility signal. The product of the error and the responsibility signals is given as the final error signal for training the forward model and the corresponding inverse model (Figure 5B). A "divide and conquer" learning strategy is implemented in this parallel architecture. A prior probability of the responsibility signal can be learned by a responsibility predictor. The responsibility predictor can be interpreted as a parallelized version of the gating network of the mixture-of-experts architecture. MOSAIC can work either in supervised (Figure 5B) or reinforcement learning (Figure 5A) paradigms (Haruno et al., 2001; Doya et al., 2002). In the case of reinforcement-learning MOSAIC, an actor or a critic or both take the role of the inverse model. Because of the dualistic nature of forward and inverse models, MOSAIC can generate a sequence of complicated behaviors, and can also recognize patterns in symbolic representations of the sequence by observing continuous movement trajectories (Kawato et al., 2000; Wolpert et al., 2003; Samejima et al., 2006; Kawato and Samejima, 2007). Based on these experimental and theoretical explorations of possible computational roles of cerebellar internal models in cognitive functions, Ito (2008) proposed a cognitive framework of cerebellar internal models for thought processes. Figure 4 shows the number of PMC papers published per year with the following three PubMed search conditions; {(internal model) AND cerebellum}, [{(internal model) AND cerebellum} AND (motor control)], [{(internal model) AND cerebellum} AND cognition]. Rapid increases in all three categories after 2000 indicate that cerebellar internal models for both motor control and cognition have begun to be widely accepted in the last two decades.

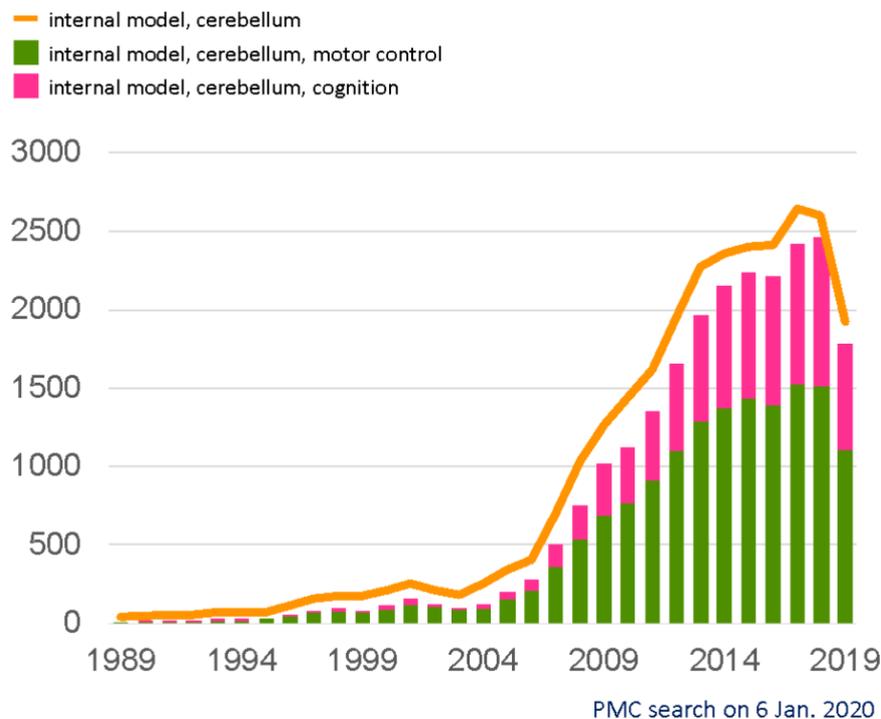

**Figure 4. Numbers of PMC papers published each year, as shown by PubMed searches for the keywords "cerebellum and internal models", and "motor control" or "cognition".** The orange line indicates [internal model AND cerebellum]. Green bars indicate [internal model AND cerebellum AND motor control], and magenta bars indicate [internal model AND cerebellum AND cognition]. The search was conducted on 6 January 2020.

## Toward a new computational theory of the cerebellum

The three models proposed by Marr, Ito, and Albus were extremely successful in establishing a motor learning framework within which our understanding of the cerebellum was advanced for oculomotor control, classical conditioning, and visually guided arm-reaching. However, these models are not sufficient to extend our understanding of cerebellar functions to whole body motions or cognition, because of computational and experimental reasons that we will discuss in this section. Furthermore, based on analyses of difficulties with the three models, we propose a new computational framework of the cerebellum.

### *The curse of dimensionality*

First, the three models do not provide insights regarding how to resolve "curse of dimensionality" associated with learning in systems with many degrees of freedom. Numbers of muscles and neurons used by animals in whole body movements are huge. Machine learning algorithms, including supervised and reinforcement learning, confront difficulties with such large systems. If a whole-body movement includes 1,000 muscles, and each muscle can generate 100 different levels of force, a simple and plain reinforcement learning algorithm needs to visit and explore on the order of $100_{1,000}$ states to find a good solution. For example, unsatisfactory results from the final competition of the DARPA robotic challenge demonstrated these difficulties even with systems having far fewer degrees of freedom than biological systems, that is, humanoid robots (Atkeson et al., 2018). Under some mathematical assumptions, a statistical learning theory estimates a generalization error for a test dataset not used for training, as $d/2n$, where $d$ is the number of parameters in the classification algorithm and $n$ is the number of training samples (Watanabe, 2009). Recent asymptotic theories motivated by the success of deep neural networks estimated that the number of necessary training samples for good generalization could be relaxed, compared with the above estimate (Zhang, 2017; Schmidt-Hieber, 2017; Suzuki, 2018; Amari, 2020), but tens of millions of training samples are still needed for tens of millions of synapses. Considerations of these theoretical constraints for cerebellar learning suggest an astronomical number of necessary training samples to achieve good generalization, with unacceptably long training periods, as follows. Let us assume that there exist 10,000 microzones in the human cerebellum, and that each microzone works as a single machine-learning entity. Then each microzone possesses 1 billion parallel-fiber-Purkinje-cell synapses (Table 2). To attain a 5% generalization error (that is, 95% correct), the above estimate requires 10 billion training samples. In extremely fast sampling, let us assume that a learner collects training data at 100 Hz (1 sample per 10 ms), 3,600 sec per hour, 8 hours per day. Under these assumptions, 347 days are necessary to collect 10 billion samples. This extremely underestimated necessary training period is already too long to be biologically realistic. One of the most difficult tasks in robotics is still bipedal locomotion in rough terrain. Human babies fall only ~3,000 times during babyhood (Adolph et al., 2012). The three models do not provide insights to explore neural mechanisms for this remarkable capability of the cerebellum and the nervous system in realizing "learning from a small sample" in sensorimotor coordination for systems with huge numbers of degrees of freedom.

### *Learning of error signals*

Second, if it is granted that climbing-fiber inputs provide error signals, and that LTD-LTP of parallel-fiber-Purkinje-cell synapses is the basis of cerebellar supervised learning, the three models do not address how and where learning of error signals originates for higher functions, including manipulation of tools and language acquisition. For phylogenetically old functions, such as the vestibulo-ocular reflex or ocular following responses, error signals carried by climbing-fibers are genetically determined from retinal slips within the accessory optic system. In contrast, tool manipulation or language acquisition is a recent event, occurring within the last several hundred thousand years of human evolution. It is implausible that climbing-fiber error signals are genetically determined for these new functions of the cerebellum. As explained in the Section, Codon theory, the sign and coordinate frame of climbing-fiber error signals should be strictly matched to those of the output from the corresponding part of the deep cerebellar nuclei, so that supervised learning works. The cerebral cortex, which most probably sends original pieces of information to shape climbing-fiber inputs for the new functions, is too far from the cerebellum to be able to align the sign and coordinates between climbing-fiber input and output from the cerebellum, that is, the cerebellar input-output consistency. The three models do not provide insights about how climbing-fiber inputs are learned in the cerebellum while satisfying cerebellar input-output consistency.

*Dimensional mismatch*
Third, the three models do not provide theoretical clues to resolve a dimensional mismatch between one degree-of-freedom climbing-fiber inputs and extremely large dimensions to be controlled in whole body movements (Morton and Bastian 2007; Hoogland et al., 2015; Machado et al., 2015) and higher cognition (Strick et al., 2009; Lu et al., 2012; Schmahmann et al., 2019). Within each microzone, climbing-fiber inputs are similar (Andersson and Oscarsson, 1978: Ito et al., 1982, Apps and Garwicz, 2005; Apps and Hawkes, 2009); thus, the error signal is basically one-dimensional. In oculomotor control, three pairs of extraocular muscles approximately determine horizontal, vertical, and rotational axes of motor coordinates. Accordingly, three microzones of the flocculus possess horizontal, vertical, and rotational retinal slips as climbing-fiber inputs (Andersson and Oscarsson, 1978: Ito et al., 1982). For oculomotor control, the three dimensions of motor coordinates are represented by three microzones of the flocculus. Thus movement and cerebellar dimensions are matched. However, the three models do not provide theoretical insights into how and where extremely large degree-of-freedom systems for whole body motion or cognition can be controlled and learned with single degree-of-freedom climbing-fiber inputs. This theoretical issue is tightly coupled to the first and second difficulties that we explained above. If drastic reduction of dimensions from for example 1,000 to only 1 is possible by some neural mechanisms while preserving the sign and coordinate frames of climbing-fiber inputs and cerebellar output so that supervised learning works efficiently, this dimensional reduction simultaneously solves three difficulties: the curse of dimensionality, learning of error signals, and dimensional mismatches between huge problems and cerebellar climbing-fiber inputs. Some recent studies propose possible roles of synchronization of inferior olive neurons by electrical synapses, and meta-cognition, in reduction of effective degree-of-freedom for learning systems to resolve the curse of dimensionality (Tokuda et al., 2009, 2013, 2017; Kawato et al., 2011; Hoang et al., 2019; Cortese et al, 2019a, b).

*Recent experimental findings that cannot readily be explained*
Recent experimental advances in cerebellar research led to findings that cannot be readily explained by the three models of Marr, Ito, and Albus. These findings include memory formation in granule cells (Yang et al. 2016; Yamada et al. 2019), temporal-difference prediction errors in climbing-fibers (Ohmae and Medina, 2015), reward-related signals in climbing-fibers (Heffley et al., 2018; Heffley and Hull, 2019; Kostadinov et al., 2019; Larry et al., 2019; Tsutsumi et al., 2019), in parallel fibers (Wagner et al., 2017), and in cerebellar output (Chabrol et al., 2019; Carta et al., 2019; Sendhilnathan et al., 2020), simple-spike dependent memory formation (Lee et al., 2015), control of Purkinje-cell plasticity mediated by molecular-layer interneurons (Rowan et al.2018), and neuronal heterogeneity between Aldolase C (zebrin) zones (Sillitoe and Joyner 2007; Zhou et al. 2014; Xiao et al. 2014; Tsutsumi et al., 2015, 2019; Tang et al. 2017). Here we postulate hierarchical reinforcement learning with multiple internal models (Kawato et al., 2000; Haruno et al., 2001, 2003; Doya et al., 2002; Sugimoto et al., 2006, 2012a, 2012 b; Kawato and Samejima, 2007) as a new computational framework of the cerebellum for resolution of the three theoretical difficulties, as well as to accommodate the above experimental findings (Figure 5).

*Biological constraints for computational exploration*
Evolutionarily, the basal ganglia emerged 560 million years ago, the cerebellum 420 million years ago, and the cerebral cortex emerged 250 million years ago. The main goal of the nervous system is to maximize fitness by selecting optimal behaviors. Thus, it is a reinforcement learning task. In specific situations, animals need to achieve appropriate sub-goals that are derived from the main goal, so as to maximize fitness. Sub-goals should include foraging for food, finding mates, and avoiding injury. A fundamental assumption of our hypothesis is that each of the basal ganglia individually, the basal ganglia and the cerebellum combined, or the basal ganglia, the cerebellum, and the cerebral cortex all combined, should be able to execute most important functions at that evolutionary stage, and to add the next most important and essential functions by development and addition of a new brain division (Northcut, 2002; Shepherd and Rowe, 2017; Naumann et al., 2015; Tosches et al., 2018). We postulate that the three brain divisions, the basal ganglia, the cerebellum, and the cerebral neocortex all have a single computational goal, in contrast to the proposal by Doya (1999). The goal is reinforcement learning to select optimal behaviors, although representations and main computational learning algorithms differ between the three divisions, as discussed in the Section, Synaptic plasticity in achieving supervised learning. The basal ganglia alone in agnathans should provide a plain reinforcement learning system for action selection, with its direct and indirect pathways differentially dependent on dopamine (Ericsson et al., 2011). The plain reinforcement learning system of agnathans is not provided with computational powers of internal models and/or working memory systems that became available only later in evolution. The cerebellum in gnathostomes (fishes with jaws) combined with the basal ganglia should be a reinforcement learning system with hierarchy, modularity, and multiple internal models (Figure 5). Equipped with 10,000 microzones, enormous numbers of granule cells, and Purkinje-cell LTD-LTP, the cerebellum is well suited for temporal processing of different inputs within several hundred milliseconds (Medina and Mauk, 2000; Ohmae et al., 2013; Ohmae et al., 2017; Kunimatsu et al., 2018; Kameda et al., 2019; Sanger and Kawato, 2020). Temporal information processing capability is maximally used in multimodal integration, forward and inverse model learning, and in reward-related information processing (Yamazaki and Lennon, 2019). The cerebral neocortex in mammals combined with the basal ganglia and the cerebellum is an even more sophisticated reinforcement learning system with additional computational powers acquired by pyramidal neuron STDP and abundant recurrent connections. Multimodal integration, associative memory, topographic representations, and cortical temporal dynamics can select abstract, concise, categorical representations of behavioral contexts for fast and efficient reinforcement learning. Those are necessary computations for adaptation and learning of nursing, social and communication behaviors.

Inputs and outputs for reinforcement learning differ in the three brain divisions. While the pyramidal tract contains 2 million-fibers, the globus pallidus internal contain 350,000 neurons (Hardman et al., 2002), and the interpositus nucleus contains only 70,000 neurons in humans (Fukutani et al., 1992; Andersen et al., 2004). Thus, the ratio of output channel bandwidths for direct motor control of the cerebral cortex, the basal ganglia, and the cerebellum is 30: 6 : 1. Consequently, the cerebellum is constrained to act on a set of abstract motor and reward related variables. This is related to the one-dimension error signal spanned by climbing-fiber inputs within a single microzone. If climbing-fiber inputs encode reward/penalty related information, corresponding parts of the cerebellum can learn state-value functions, state-action value functions, reward prediction errors and other reward-related variables from parallel-fiber inputs. The cerebrocerebellum is connected to the cerebral cortex via both inputs and outputs, and it can utilize efficient, concise spatiotemporal representations acquired in the cerebral cortex (Gao et al., 2018; Wagner et al., 2019). It was recently revealed that for some parts, an output from one microzone is fed to granule cells of another microzone (Houck and Person, 2014; Gao et al., 2016; Giovannucci et al., 2017), implying that these microzones are hierarchically/heterarchically arranged (Ohmae S, private communication) (Figure 5B). Thus, we have multiple modules, internal models, and hierarchy/heterarchy in the cerebellum.

*Hierarchical and modular reinforcement learning in the cerebellum*
Hierarchical and/or heterarchical reinforcement learning frameworks with multiple internal models (Kawato et al., 2000; Haruno and Kawato, 2006) could provide theoretical guidelines to resolve the three computational difficulties of the models of Marr, Ito and Albus as well as to interpret new experimental findings. The three computational difficulties of the three models are the curse of dimensionality, learning of error signals, and dimensional mismatch. The proposed frameworks resolve the curse of dimensionality with hierarchical structure because the top layer can search for the optimal solution in a much lower dimension thanks to the dimensional reduction by the bottom layer and requires only a moderate number of learning trials. Multiple models connected serially rather than in parallel provide a computational architecture where one module can learn the error signal, which is used in learning by the other module. Hierarchical and/or heterarchical structures could connect top and bottom layers with huge differences in dimensions. Here, the top layer may possess very low-dimensional symbolic representations, while the bottom layer may possess extremely high-dimensional representations related to motoneurons and muscles. We note that the three models did not discuss hierarchy, heterarchy, serially connected multiple modules, or how representations with huge differences in dimensionality could be related within the cerebellum. New frameworks may provide several clues to interpret the new experimental findings, and are partially motivated by biological constraints including evolutionary views about the brain.

The hierarchical reinforcement learning scheme with multiple modules shown in Figure 5A (Sugimoto et al., 2006) illustrates how abstract symbolic representations in the top layer are implemented as responsibility signals in the bottom layer. This model is a combination of a reinforcement MOSAIC (Doya et al., 2002) and hierarchical MOSAIC (Haruno et al., 2003). In each layer, reinforcement learning occurs with multiple modules. The responsibility signal determines which forward model in the bottom layer best represents the environment. Patterns of responsibility signals allow the hierarchical MOSAIC to achieve discretization of continuous space. Changes of dynamics in the continuous space are approximated by Markov transitions between the abstract and discrete states. The top layer controls this transition as well as providing sub-goals to the bottom layer. Drastic dimension reduction is achieved by this abstraction in the top layer. In the bottom layer, a "divide and conquer" strategy works with many experts (forward models), even for highly nonlinear problems. Abstraction and divide-and-conquer together resolve the curse of dimensionality. In most successful engineering applications of hierarchical reinforcement learning algorithms in robotics, higher-level abstract representations and/or sub-goals of the bottom layer were determined by researchers (Atkeson et al., 2000). So far, intermediate goal postures for a standing robot were manually selected as representations in the top layer by Morimoto and Doya (2001). As another example, Bentibegna et al. (2003) selected right-bank shots as higher-level actions in air-hockey by a humanoid robot DB. Neither the cerebellum nor the whole brain can enjoy the luxurious special treatment that a homunculus in the brain selects appropriate representations for the top layers. From a neuroscience point of view, abstract representations of movements in the top layers could be movement sub-goals, synergies, motion of center of gravity, locomotion gates, periods and phases of rhythmic movements, and phase relationships of limb rhythms. Computationally, the most difficult question is how appropriate high-level representations can be automatically acquired without homunculus or genetic hard-wiring. Hierarchical and reinforcement MOSAIC can find higher-level representations through the pattern of responsibility signals, that is, which combination of forward models, value functions and policies is most appropriate in a given situation, and could provide a computational possibility for automatically learning sub-goals and synergy.

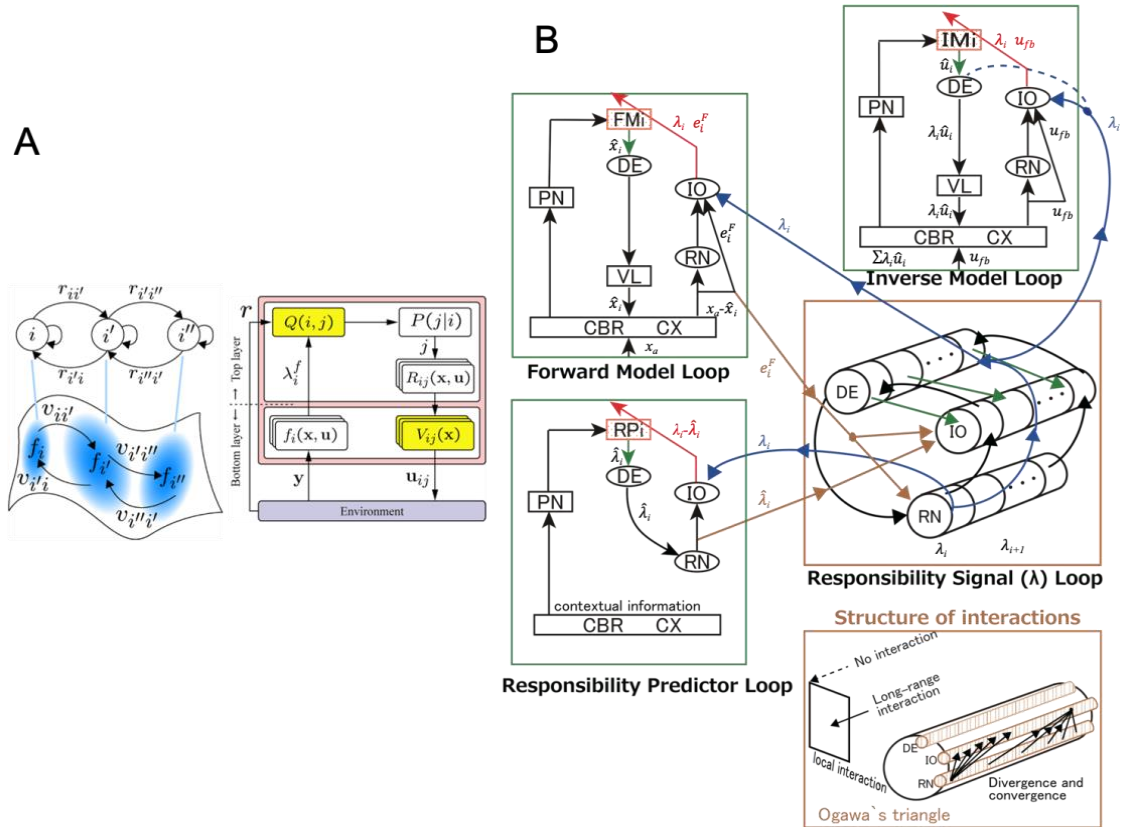

**Figure 5. A hierarchical reinforcement learning model with multiple modules. A.** A hierarchical reinforcement learning model with top and bottom layers based on reinforcement MOSAIC (Doya et al., 2002) and hierarchical MOSAIC (Haruno et al., 2003) (adapted from Sugimoto et al., 2006). **(left panel)** The lower curved sheet represents continuous state-*x* space of the environment with nonlinear dynamics. The continuous space is automatically divided into discrete regions (blue shades) according to goodness of prediction by each forward model, while the MOSAIC architecture predicts the whole nonlinear dynamics with different combinations of forward models, $f_i$, where the index *i* is given for each forward model. This discrete division of the continuous state-space allows this MOSAIC to treat the transitions between state-space patches (blue regions) as discrete state transitions in semi-Markov decision processes (*i, i', i''* in top layer). **(right panel)** The top layer maximizes a reward *r* provided by the environment by controlling state-*i* transition while selecting the abstract action-*j*, and influencing bottom layer reinforcement learning by providing inter-layer rewards $R_{ij}(x,u)$ for the state *i* and the action *j*. In other words, the top layer supplies sub-goal information to the bottom layer. The whole hierarchical reinforcement learning is driven by the reward *r*, while inter-layer communication

was achieved as a state abstraction by forward models in the bottom layer, and setting sub-goals for the bottom layer by the top layer. The top layer symbolically represents which reinforcement-learning modules in the bottom layer are selected, and controls Markov transitions between the discrete states such as *i, i', i''*, while selecting the abstract action *j* that determines inter-layer rewards for the bottom layer. Reinforcement Q-learning takes place with the state-action value function, *Q(i,j)*. $P(j|i)$ denotes *j-th* action selection probability for the *i-th* state, i.e., top-layer policy, and is determined from *Q(i,j)*. The bottom layer is a continuous reinforcement learning system to learn an action *u* such that $R_{ij}(x,u)$ is maximized while the top layer is in the *i-th* state and takes the action *j*. When learning is achieved, the top layer maximizes the reward in much smaller dimensions (discrete space of *i*) than in the original large dimensions of the environment, and the bottom layer maximizes the sub-goals (inter-layer rewards) from the top layer so that the bottom layer faithfully transforms the actions of the top layer in the discrete space into the control of *x* in the continuous space. $V_{ij}(x)$ is the value function, which is learned from the reward $R_{ij}(x,u)$. *x*, bottom-layer state, *y*, bottom-layer observation, *u*, bottom-layer action, $f_i(x,u)$, *i*-th forward model, $\lambda_i^f$, responsibility signal of the *i*-th forward model, $u_{ij}$ is the bottom-layer action under the *i-th* region with value function $V_{ij}(x)$, and the top-layer state *i* and action *j*. Note that there are no inverse models or responsibility predictors in this hierarchical reinforcement MOSAIC. **B.** Another MOSAIC model with loop structures implements interactions between the three modules for the inverse model, forward model, and responsibility predictor (adapted from Kawato et al., 2000). $FM_i$, $IM_i$, and $RP_i$ indicate the *i*-th forward model, the *i*-th inverse model, and the *i*-th responsibility predictor, respectively, realized by Purkinje cells. $\lambda_i$ is the *i*-th responsibility signal of the *i*-th modules, which is common to $FM_i$, $IM_i$, and $RP_i$. $x$ is an actual state. $\hat{x}_i$ is a predicted state by $FM_i$. $e_i^F = x - \hat{x}_i$ is the prediction error signal of the state, and $\lambda_i e_i^F$ is the final error signal used to change $FM_i$ in supervised learning. $\hat{u}_i$ is the *i*-th motor command generated by $IM_i$. $\sum_i \lambda_i \hat{u}_i$ is the responsibility-signal weighted summation of the motor commands from all inverse models, and this is the total motor command output from the cerebellum. $u_{fb}$ is a feedback motor command. $\lambda_i u_{fb}$ is the final error signal to change $IM_i$ used in supervised learning. $\hat{\lambda}_i$ is the *i*-th predicted responsibility signal computed by $RP_i$. $\lambda_i - \hat{\lambda}_i$ is the error in responsibility prediction to change $RP_i$. Red arrows indicate the three different kinds of error signals that drive supervised learning of the three modules. Blue arrows show responsibility signals that control common weighting in supervised learning of the three modules. Brown arrows show prior (predicted responsibility) and likelihood for responsibility estimation. Predicted responsibility signals are computed before movement execution. The three kinds of modules interact with each other in feedback connections between the cerebellar cortex, the dentate nucleus (DE), the red nucleus (RN), and the inferior olive nucleus (IO). Convergent and divergent connections between different modules anatomically implement hierarchical/heterarchical structures between different layers, and are shown in the triangle consisting of DE, RN and IO. VL, ventrolateral thalamus, CBR CX, cerebral cortex, PN, pontine nucleus.
*Figure **A** was published in The IEICE transactions on information and systems (Japanese edition), 89(7), Sugimoto N, Samejima K, Doya K, Kawato M, Hierarchical Reinforcement Learning: Temporal Abstraction Based on MOSAIC Model, 1577-1587, Copyright IEICE (2006).
*Figure **B** was published in Kagaku, 70(11), Kawato M, Doya K, Haruno M, Multiple paired forward and inverse models (MOSAIC) – the information-processing and possibility, 1009-1017, Copyright Iwanami Shoten (2000).

The heterarchical MOSAIC shown in Figure 5B (Kawato et al., 2000) shows how the three modules, forward models, inverse models, and responsibility predictors, are heterarchically arranged by convergent and divergent connections within the closed circuit consisting of the dentate, red, and inferior olive nuclei. Responsibility predictors can predict responsibility signals, which estimate goodness of prediction of each forward model from pieces of contextual information. In Allen and Tsukahara (1974) and Kawato and Gomi (1992), functions of the closed loop circuit between the dentate nucleus, the parvo-cellular part of the red nucleus, and the inferior olive sending climbing-fibers (De Zeeuw et al., 1998; Sokolov et al., 2018) were discussed, but remain enigmatic, although this Ogawa's triangle (also known as the triangle of Guililain and Mollaret) is huge in humans (Ogawa, 1941). A recent study reported that cerebellar output during eyeblink conditioning can contribute to generation of an error signal in climbing-fibers (Ohmae and Medina 2019). If this triangle is a neural connection from outputs of one microzone to error signals of another, the second theoretical difficulty can be tackled in the hierarchical reinforcement learning framework. In the upper level of hierarchical reinforcement learning, abstract, concise representations could be used and the curse of dimensionality would be drastically relaxed. With the cascade arrangements of microzones, climbing-fiber error signals for one microzone could be learned by the other upper-layer microzone. Hierarchical structures could bridge concise, abstract representations in the top layer with high-dimensional representations in the bottom layer. One-dimensional error signals for each module in each layer could provide a mechanism to find a low-dimensional order parameter for motor control represented in its outputs.

## Conclusion

This is a new era of cerebellar research with veritable explosions of new techniques, new findings, and new computational ideas. It is important and timely to evaluate contributions and limitations of the Marr, Ito, and Albus models and to explore future directions based on advances made following their proposals. In this review, we have made a computationally oriented effort, and in introducing these models we have experienced the same excitement that we felt when we first encountered these theories 50 years ago.

**Acknowledgements** We thank Drs. Keisuke Toyama, Tomoo Hirano, Soichi Nagao, and Kazuo Kitamura for their comments on an earlier version of the manuscript. This research was partially supported by AMED under Grant Number JP19dm0307008. MK and HH were further supported by JST ERATO (Japan, grant number JPMJER1801).